\documentclass{aa}  
\usepackage{graphicx}
\usepackage{natbib}
\usepackage{txfonts}

\begin{document} 

   \title{Discovery of a jet-like structure with overionized plasma in the SNR IC443}

   \author{Emanuele Greco
          \inst{1,2}
          'and Marco Miceli\inst{1,2} \and Salvatore Orlando\inst{2} \and Giovanni Peres\inst{1,2} \and Eleonora Troja\inst{3,4} \and Fabrizio Bocchino\inst{2}
	  }

   \institute{Dipartimento di Fisica \& Chimica, Universit\'a  di Palermo, Piazza del
Parlamento 1, 90134 Palermo, Italy
         \and
             INAF-Osservatorio Astronomico di Palermo, Piazza del Parlamento 1,
             90134, Palermo, Italy
         \and
             NASA Goddard Space Flight Center, 8800 Greenbelt RD, Greenbelt, Maryland 20771, USA
         \and
             Department of Astronomy, University of Martland, College Park, MD 20742-4111, USA.
             }

  \date{Sent January 31, 2018; accepted April 12, 2018}

\abstract{\\ \emph{Context.} IC443 is a supernova remnant located in a quite complex
environment since it interacts with nearby clouds. Indications for the presence of
overionized plasma have been found though the possible physical causes of
overionization are still debated. Moreover, because of its peculiar position and
proper motion, it is not clear if the pulsar wind nebula (PWN) within the remnant is
the relic of the IC443 progenitor star or just a rambling one seen in projection on
the remnant.  \\ \emph{Aims.} Here we address the study of IC443 plasma in order to
clarify the relationship PWN-remnant, the presence of overionization and the origin
of the latter. \\ \emph{Methods.} We analyzed two \emph{XMM-Newton} observations
producing background-subtracted, vignetting-corrected and mosaicked images in two
different energy bands and we performed a spatially resolved spectral analysis of
the X-ray emission. \\ \emph{Results.} We identified an elongated (jet-like)
structure with Mg-rich plasma in overionization. The head of the jet is interacting
with a molecular cloud and the jet is aligned with the position of the PWN at the
instant of the supernova explosion. Interestingly, the direction of the jet of
ejecta is somehow consistent with the direction of the PWN jet.\\
\emph{Conclusions.} Our discovery of a jet of ejecta in IC443 enlarge the sample of
core-collapse SNRs with collimated ejecta structures. IC443's jet is the first one
which shows overionized plasma, possibly associated with the adiabatic expansion of
ejecta. The match between the jet's direction and the original position of the PWN
strongly supports the association between the neutron star and IC443.\\

}
\titlerunning{Overionized jet in SNR IC443}
\authorrunning{Greco et al.}
\maketitle

\maketitle

\section{Introduction}

Supernova Remnants (SNRs) are divided into different categories according to different morphological features. One of the most interesting and less understood is the \emph{Mixed Morphology} (MM) class which shows thermal peaked X-ray emission in the inner shells and peaked radio emission in the remnant outline. In the last years, several works found signatures of overionized plasma in this class of SNRs (e.g. \citealt{oky09}, \citealt{mbd10}, \citealt{zmb11} for W49b; \citealt{uky12} and \citealt{smi14} for W44; \citealt{sk12} for W28). For IC443, marginal evidence of overionized plasma was detected in the inner region characterized by bright X-ray emission (\citealt{kon02}, \citealt{tbm08}), and later confirmed by Suzaku observations (\citealt{yok09}, \citealt{out14} and \citealt{mtu17}, hereafter M17). 

 IC443 (also called G189.1+3.0) belongs to the GEM OB1 association at a distance of 1.5 kpc (\citealt{ws03}). Its radius is $\approx$ 20', its age is $\approx 4000$ yr (\citealt{tbm08}) and it is a MM SNR (\citealt{rp98}). It is located in a very complex environment since it interacts with a molecular cloud in the northwestern (NW) and southeastern (SE) areas and with an atomic cloud in the northeast (NE). The molecular cloud lies on the foreground of IC443 forming a semi-toroidal structure (\citealt{bgb88}, \citealt{tbr06}, \citealt{sfy14}).

The PWN CXOU J061705.3+222127 is observed within the remnant shell. However, since it is far away from the geometric center of the remnant (it is near the southern edge) and moves towards southwest (SW), it is not clear if the PWN belongs to IC443 (\citealt{gcs06}, \citealt{spc15}) or is a rambling neutron star (NS) seen in projection on the remnant. 

In this paper we present our analysis of deep XMM-Newton observations of IC443, which revealed for the first time a jet-like structure of overionized plasma in the NW part of the remnant. The identification of this structure is interesting because X-ray emitting jets of ejecta have been discovered and analyzed only in other two core-collapse SNRs, namely Cas A (\citealt{hlb04}, \citealt{fhm06}) and, recently, Vela SNR (\citealt{gsm17}). This kind of structures are not well understood and some authors suggested that they may be deeply related to the physical mechanism of the SN explosion (\citealt{wil85}, \citealt{jms16}, \citealt{gs17}).

 The paper is organized as follows: observations and data reduction are described in Sect. \ref{obs}; image and spectral analysis are illustrated in Sect. \ref{res}; discussion about results can be found in Sect. \ref{discu}.
\section{Observations and data analysis}
\label{obs}

IC443 has been observed several times with the European Photon Imaging Camera (EPIC) on board of XMM-Newton. We analyzed nine observations in order to build complete images of IC443 but only two of them (ID 0114100201 and 0114100501, PI Fred Jansen) were used for the spectral analysis since they are the only ones which have the jet within the field of view (FOV) (Table \ref{osservazioni}). 

\begin{table}[!h]
\centering
\begin{tabular}{c|c|c|c}
\hline\hline
OBS ID& Camera& $t_{exp}$ U (ks)& $t_{exp}$ F (ks)\\
\hline
0114100101& MOS1& 22.9 & 9.5 \\ 
0114100101& MOS2& 22.9 & 9.5 \\
0114100101& PN& 26.7 & 3.7 \\
\hline
0114100201& MOS1& 5.4& 5\\ 
0114100201& MOS2 &5.4& 5.2\\
0114100201& PN& 3& 2.6\\
\hline
0114100301& MOS1& 25.9 & 21.3 \\
0114100301& MOS2& 25.6& 21.5 \\
0114100301& PN& 23.2 & 17.7 \\
\hline
0114100401& MOS1& 29.9 & 23.5\\
0114100401& MOS2& 29.9 & 23.8\\
0114100401& PN& 27.8 & 18.7\\
\hline
0114100501& MOS1& 24.9& 18.1\\
0114100501& MOS2& 24.9& 18.9\\
0114100501& PN& 22.5& 12.8\\
\hline
0114100601& MOS1& 6.3 & 4.8\\
0114100601& MOS2& 6.3 & 5.3\\
0114100601& PN& 3.9 & 3.2\\
\hline
0301960101& MOS1& 81.5& 56.3\\
0301960101& MOS2& 81.5& 58.9\\
0301960101& PN& 77.4& 51.2\\
\hline
0600110101& MOS1& 90.2 & 32.3\\
0600110101& MOS2& 90.3 & 37.5\\
0600110101& PN& 84.4 & 22.2\\
\end{tabular}
\caption{Main information about all observations. U stands for unfiltered, F for filtered}
\label{osservazioni}
\end{table}

We used the Science Analysis System (SAS), version 16.1.0, to perform the whole data analysis. We used the task \emph{ESPFILT} to eliminate contamination due to soft protons, reducing the effective exposure time (third column of Table \ref{osservazioni}). The task ESPFILT extracts two lightcurves for the events detected inside and outside of the FOV and creates a high energy count rate histogram from the field-of-view data. The histogram has a Gaussian peak at the nominal count rate (i.e. the count rate measured during time intervals not contaminated by protonic flares) with a high count rate tail. The shape of this Gaussian peak significantly depends on the data contamination. If the peak is narrow, ESPFILT is not able to properly filter events. However, in our case, the peak has a bump-like shape and ESPFILT can fit a Gaussian to the peak and determine thresholds that we set at 1.5$\sigma$. Time intervals corresponding to the points within this threshold are labelled as "good time intervals", while those above it are removed from the event list. For the whole analysis we considered only data with FLAG=0 and PATTERN$<$13,$<$5 for MOS,PN cameras respectively, as the XMM-Newton handbook suggests\footnote{$https://xmm-tools.cosmos.esa.int/external/xmm-user-support/documentation/uhb/$}. 

We built background-subtracted images of IC443 in two different bands: Soft (0.5-1.4 keV) and Hard (1.4-5.0 keV), same as in \cite{tbm08}. We used a double subtraction procedure in order to subtract background contribution and correct for the vignetting, similar to \cite{mbo17}. We subtracted the non-photonic background in our data and in the EPIC high signal-to-noise background event files\footnote{Such files can be found at XMM ESAC website https://www.cosmos.esa.int/web/xmm-newton/blank-sky} by scaling the EPIC Filter Wheel Closed files\footnote{Such files can be found at XMM ESAC website $https://www.cosmos.esa.int/web/xmm-newton/filter-closed$} data. Before subtraction, we used the ratio of count-rates in the corners of the CCDs, namely the part of the cameras outside of FOV, as a scaling factor. Then, we subtracted the photonic background from our pure photonic data. Moreover, we combined the observations using the task \emph{emosaic}. Finally, we adaptively smoothed the mosaicked image with the task \emph{asmooth} to a signal to noise ratio of 25. In all images, North is up and East is to the left.

We used the SAS tool \emph{evigweight} to correct vignetting effect in the spectra. We applied tasks \emph{rmfgen} and \emph{arfgen} obtaining response and ancillary matrices and we binned spectra to obtain at least 25 counts per bin. The spectral analysis has been performed with XSPEC (version 12.9, \citealt{arn96}) in the energy range 0.6-5 keV for all cameras. PN and MOS1,2 spectra of the two observations were fitted simultaneously. We selected two different regions outside the shell to extract background spectra and we verified that the best fit results do not depend on the choice of the background. Errors are at 2$\sigma$ confidence levels and $\chi^2$ statistics are used.

We also built a \emph{Chandra} image of the PWN CXOU J061705.3+222127 located near the southern rim of the remnant in the 0.5-7 keV energy band. We analyzed observations 13736 and 14385 (PI Weisskopf), for a total exposure time of 152 ks, through the software CIAO version 4.7. We used the CIAO task \emph{fluximage} which created exposure-corrected images for the observations (merged with the task \emph{merge\_obs}).

Scales in pc shown in the images are calculated assuming a distance of 1.5 kpc, according to \cite{ws03}.

\section{Results} 
\label{res}

\subsection{Image analysis}
\label{image}

\begin{figure}[!ht]
\centering
\begin{minipage}{0.5\textwidth}
 \includegraphics[width=\textwidth]{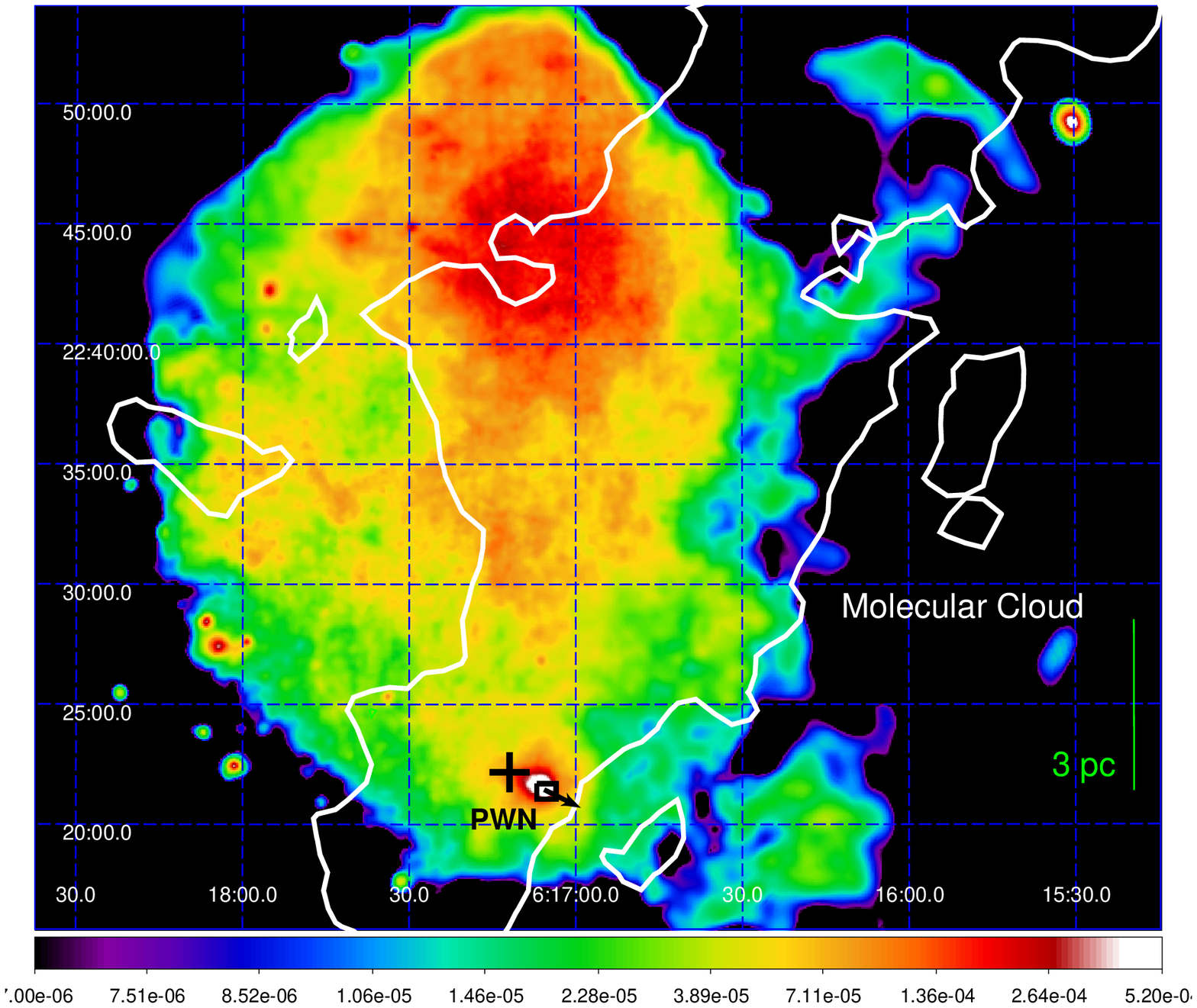}
 \end{minipage}
 \hfill
 \begin{minipage}{0.5\textwidth}
 \includegraphics[width=\textwidth]{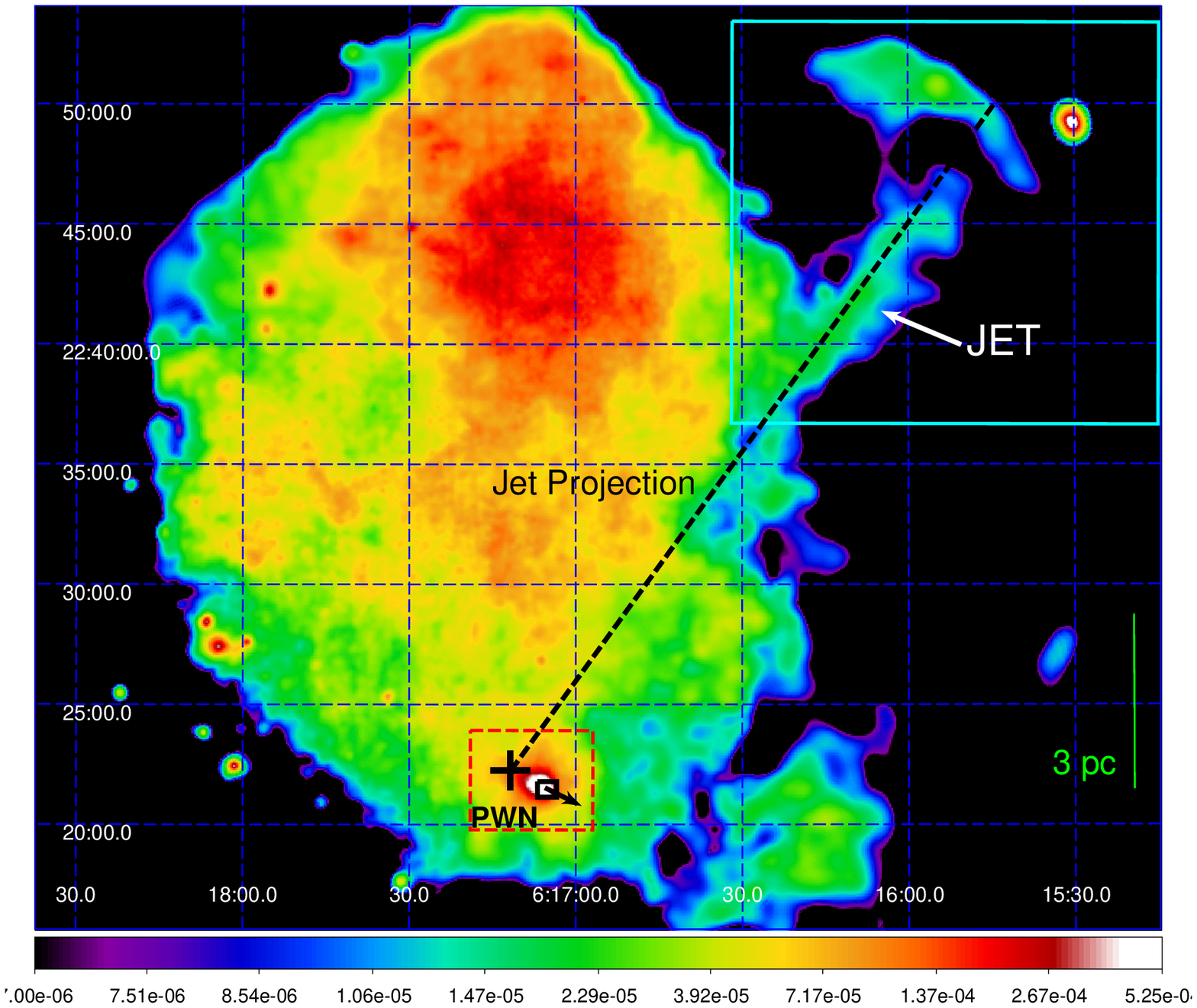}\hspace{-0.2cm}\llap{\raisebox{0.8cm}{\includegraphics[height=3.cm]{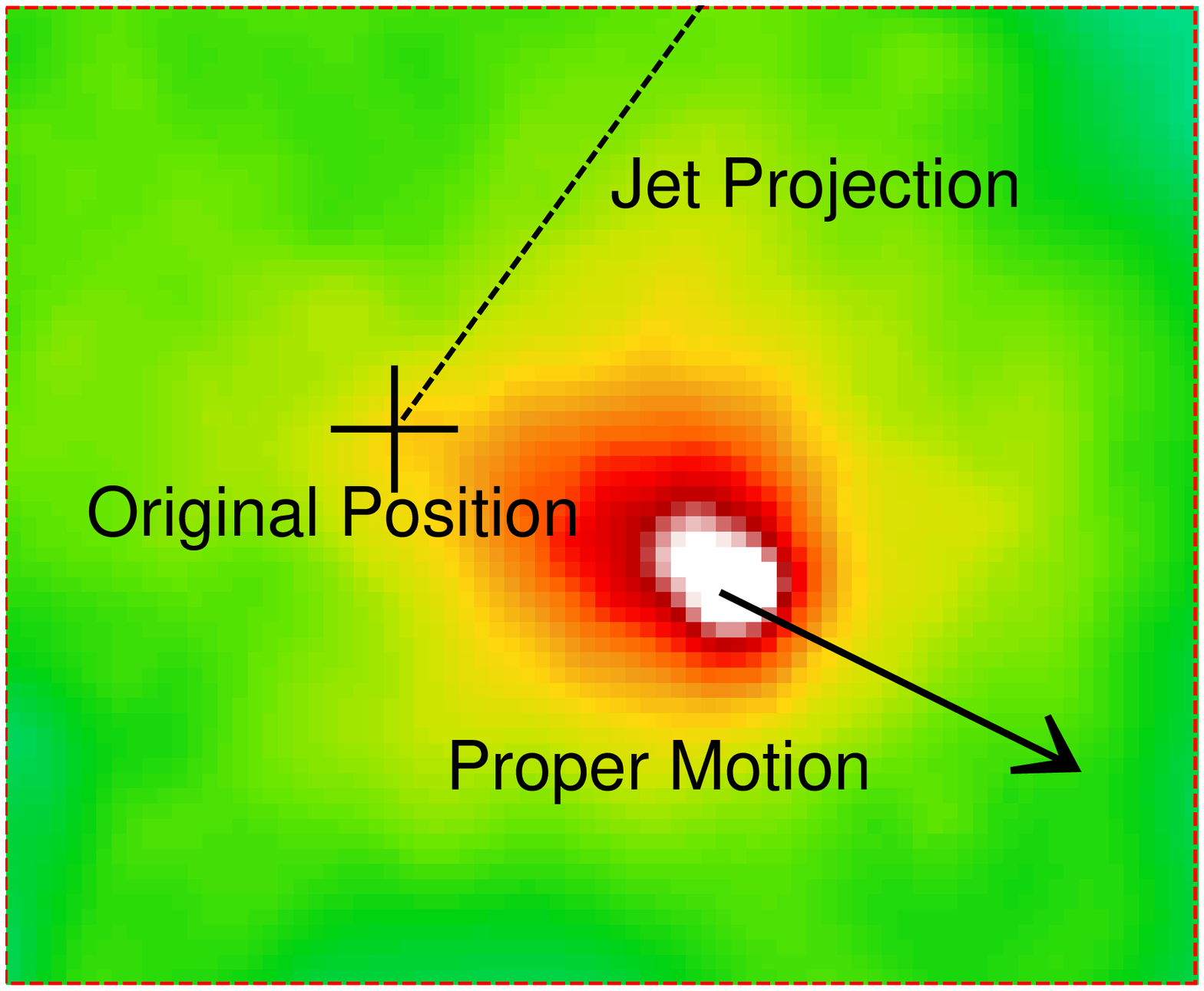}}}
 \end{minipage}
 \caption{Background-subtracted and vignetting-corrected count-rate images of IC443 in the Hard band (1.4-5 keV) obtained by combining all the EPIC cameras. The image is adaptively smoothed to a signal-to-noise ratio 25, the bin size is $6''$ and the color scale is logarithmic. In the upper panel, the black box marks the PWN (and the field of view of Fig. \ref{PWN}), the black arrows marks the proper motion of the PWN, the black cross indicates the original position of the PWN (immediately after the explosion) and the white contours traces the molecular cloud as reported by \cite{bgb88}. In the lower panel, the cyan box marks the jet (and the field of view of Fig. \ref{dettagliojet}), the dashed black line is the projection of the jet towards the PWN and the dashed red box is the FOV of the insert in the lower panel. This close-up view shows a detail of the PWN area with the original position and proper motion marked by the black cross and black arrow, respectively.}
\label{ic443andjet}
\end{figure}

\begin{figure}[!ht] 
\centering
\includegraphics[width=\columnwidth]{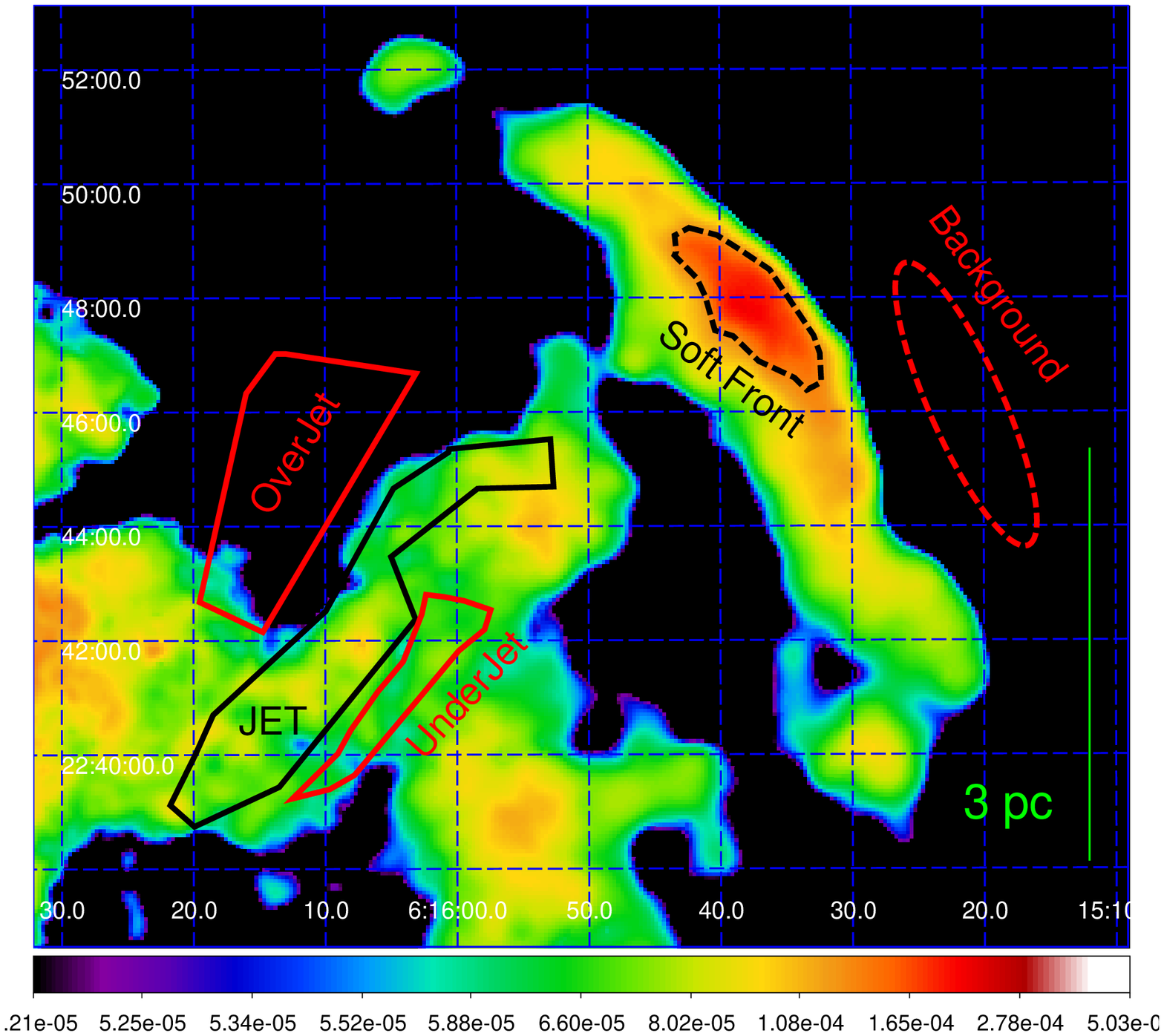}
\includegraphics[width=\columnwidth]{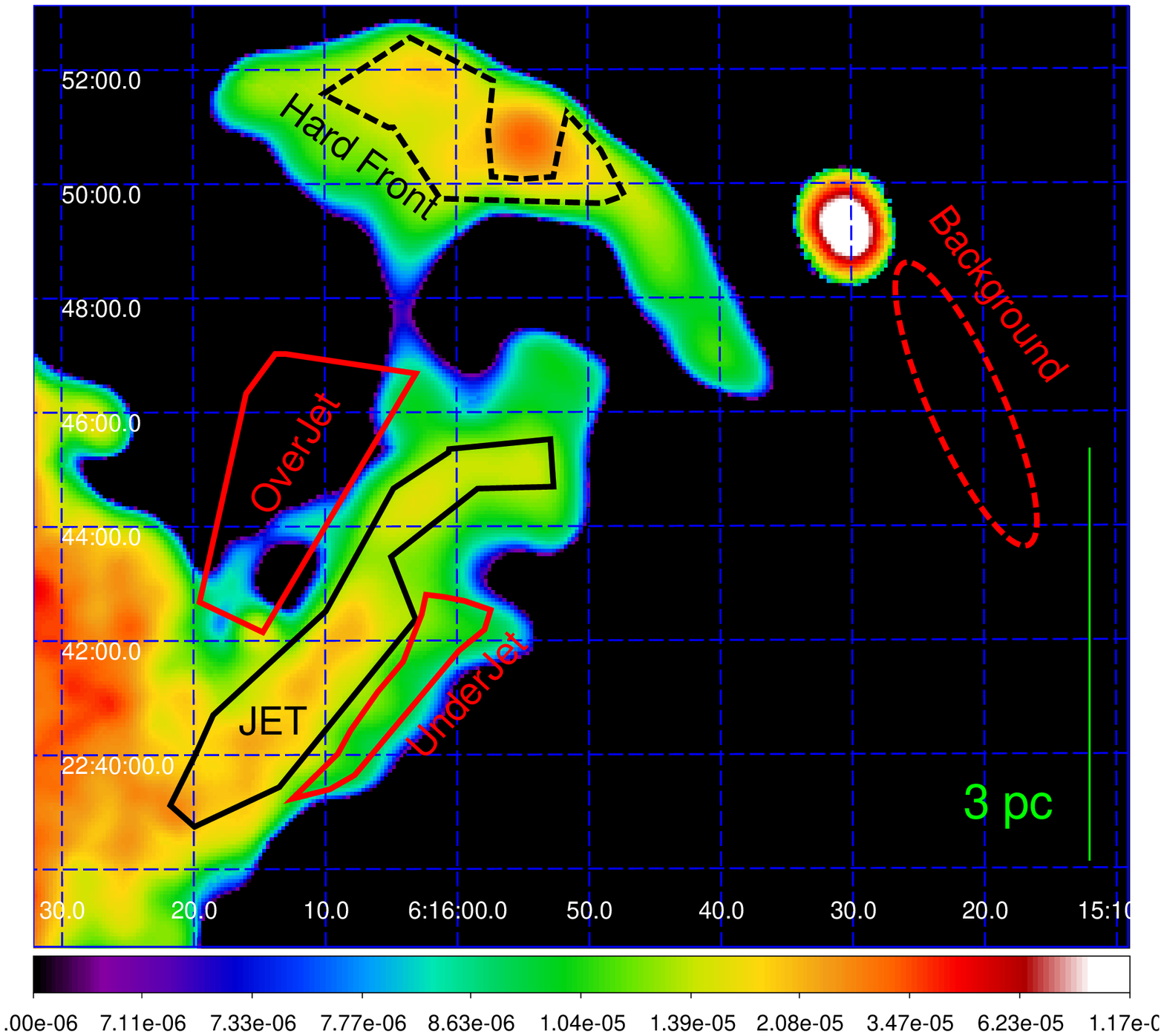}
\caption{Close-up view of the jet area marked by the cyan box in Fig. \ref{ic443andjet} in the Soft band (0.5-1.4 keV, top) and in the Hard band (1.4-5 keV, bottom). The image is adaptively smoothed to a signal-to-noise ratio 25, the bin size is $6''$ and the color scale is logarithmic. Jet region is marked in black, Soft Front and Hard Front are marked in dashed black, background region in dashed red and OverJet and UnderJet regions in red.}
\label{dettagliojet}
\end{figure}

\begin{figure}[!h]
\centering
\includegraphics[width=\columnwidth]{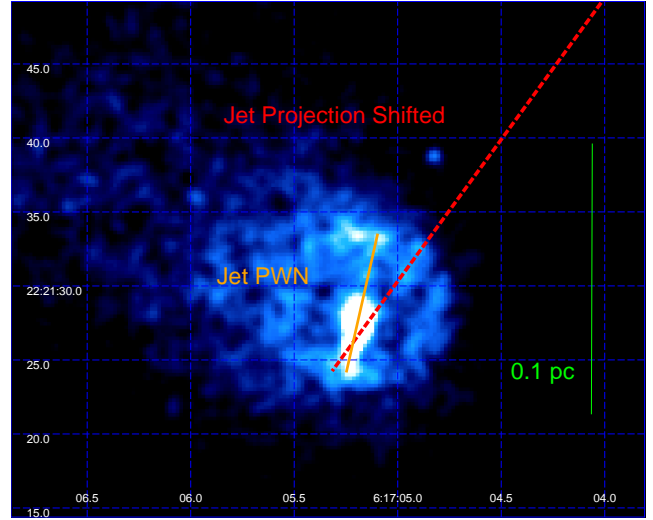}
\caption{Chandra image of the area near the PWN which provides a comparison between the SNR and the PWN jets in the 0.5-7 keV energy band.  The image is smoothed with a Gaussian with $\sigma=2$ pixel, the pixel size is $0.246''$ and the color scale is linear. The projection of the jet has been shifted on the PWN in order to compare the two jets' direction.}
\label{PWN}
\end{figure}

In Fig. \ref{ic443andjet} we display the count-rate image of IC443 in the Hard energy band (1.4-5 keV). The white contour in the upper panel traces the semi-toroidal molecular cloud mentioned above (\citealt{bgb88}). Until now, only the internal area of IC443 has been studied in detail with XMM-Newton (\citealt{tbr06}), while the elongated jet-like structure clearly evident within the cyan box (lower panel) in the NW has not been investigated yet. If we take a closer look to this area (Fig. \ref{dettagliojet}), we can see that the jet, $\approx$ 3 pc long and indicated by the central black region, is the brightest area in the Hard energy band. The jet is characterized by a relatively narrow, well-collimated morphology which is also visible in the Soft band, though with a lower contrast. If we take into account the NS proper motion ($\approx$ 130 km/s towards SW, \citealt{gcs06}) and we multiply it by the age of remnant ($\approx$ 4000 yr, \citealt{tbm08}), we can derive the position of the neutron star immediately after the explosion, as shown by the black cross in the lower panel of Fig. \ref{ic443andjet}. We can see that the projection of the jet towards the PWN crosses the position which had the NS at the time of the explosion. This result is a clear indication of the link between the PWN and IC443, given that the jet is made of ejecta expelled by the IC443 progenitor, as we will show in Sect. \ref{spectra}. We also built an image of the area near the PWN using data from the Chandra telescope (Fig. \ref{PWN}). This image was produced by rebinning the events files to $0.246''$, namely the half of the nominal pixel size ($0.492''$), as in \cite{spc15}.

 The projection of the jet is shifted on the PWN maintaining its direction in order to compare the jet's direction with the direction of the PWN jet, detected by \cite{spc15}. The directions of the two jets are somehow consistent and this may be suggestive of some physical link between their origins. In Vela SNR, the jet-like structure detected by \cite{gsm17} is not clear as in IC443 but it is still possible to measure the angle between the PWN jet and the SNR jet. In that SNR the angle is of $\approx$ 90$^\circ$, as can be derived by comparing the jet detected by \cite{gsm17} with the PWN jet showed by \cite{ptk03}.

\subsection{Spatially resolved spectral analysis}
\label{spectra}
 
\begin{figure}[!h]
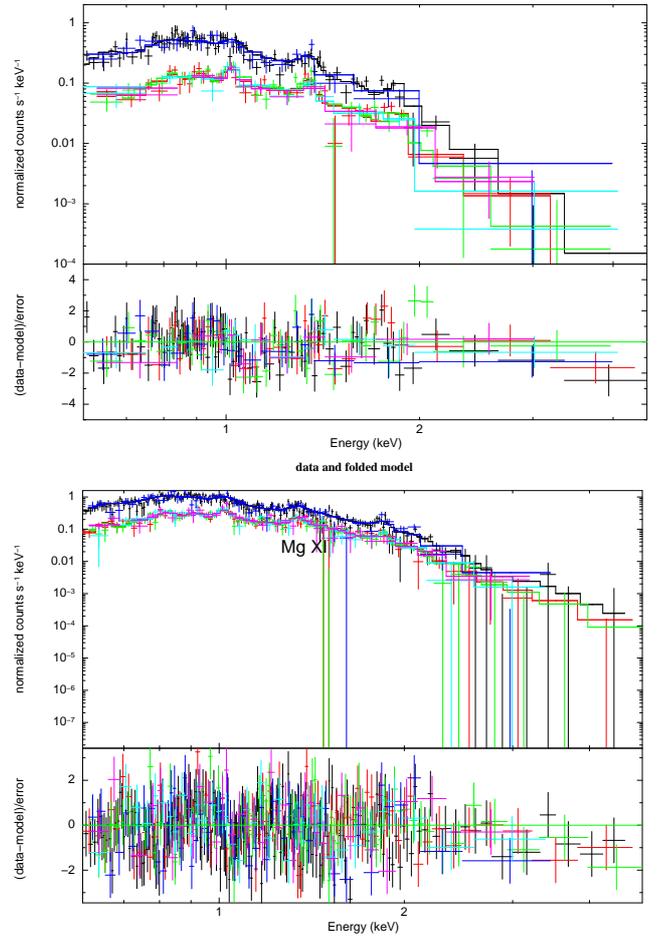

\centering
\includegraphics[angle=270,width=\columnwidth]{OverJet.ps}
\includegraphics[angle=270,scale=0.345]{SpettroJet.ps}
\caption{In the upper panel, EPIC spectra of the OverJet region obtained simultaneously fitting spectra from six cameras with a VAPEC model (best fit parameters can be found in Table \ref{spettrioverundersoft}). In the lower panel, same as is the upper one but for the Jet region described with a VAPEC+VRNEI model (best fit parameters can be found in Table \ref{risultatifit}).}
\label{spettritot}
\end{figure}

In Fig. \ref{dettagliojet} it is clear that the jet is much brighter than the adjacent regions, labelled OverJet and UnderJet (in red), particularly in the Hard band. We investigated the nature of the plasma through spectral fits and found that regions OverJet and UnderJet are both satisfyingly described by a model of plasma in collisional ionization equilibrium (VAPEC model, the spectrum relative to the OverJet region is shown in Fig. \ref{spettritot}). The resulting best-fit parameters for the two regions are all consistent within 2 sigmas, except for nH, (namely the column density $nH= \int n dl$ where n is the particle density and the integral is calculated along the line of sight). The latter result is not surprising since the molecular cloud (indicated by the white contours in Fig. \ref{ic443andjet}) is on the foreground and the OverJet region is completely covered by the cloud whereas the UnderJet region is only partially covered by it and it is close to the cloud border in projection. Since both regions are well described with a VAPEC model, it is natural to affirm that X-ray emission is associated with shocked ISM in OverJet and UnderJet.

\begin{table}[!h]
\centering
\begin{tabular}{c|c|c|c}
\hline\hline
Parameter & \multicolumn{3}{c}{Region}\\
\hline
& OverJet& UnderJet& Soft Front \\
\hline
$nH(10^{22}$ cm$^{-2})$&0.65$^{+0.04}_{-0.07}$&0.50$^{+0.06}_{-0.07}$&0.57 $ \pm 0.04$\\
\hline
kT(keV)& 0.39$^{+0.05}_{-0.01}$&0.33$^{+0.02}_{-0.01}$&0.26$\pm 0.01$\\
O&2.0$^{+0.4}_{-0.3}$& 1.4$\pm 0.3$&0.78$^{+0.11}_{-0.10}$\\ 
Mg&0.7 $\pm$ 0.1& 0.7$\pm0.1$&0.6$^{+0.1}_{-0.2}$\\
Si&0.8 $\pm$ 0.2& 0.9$ \pm0.3$&1.1$ \pm 0.5 $\\
Fe&0.29$^{+0.07}_{-0.05}$&0.42$^{+0.09}_{-0.07}$&0.41$^{+0.06}_{-0.07}$\\
\hline
$n^2 l^{*}$($10^{18}$cm$^{-2}$)&2.4$^{+0.3}_{-0.4}$&2.4$^{+0.6}_{-0.3}$&6$^{+2}_{-1}$\\
\hline
Counts&12000 & 13000 & 14000\\
$\chi^2$(d.o.f)&400.24(357)&412.79(367)&366.82(345)\\
\hline
\multicolumn{4}{c}{$^*$ Emission measure per unit area}\\
\end{tabular}
\caption{Best fit results for OverJet, UnderJet and Soft Front. S abundance is fixed equal to the Si one.}
\label{spettrioverundersoft}
\end{table}

However, when fitting the jet's spectra, we need to add a second thermal component to properly fit the spectra because a single VAPEC component gives a $\chi^2 = 643.88$ with 506 d.o.f. The best fit parameters of the soft thermal component are consistent within 2 $\sigma$ with those of the UnderJet and OverJet regions. We then fixed the abundances of this soft component to those of the OverJet region and we forced the temperature of this component to vary in the range 0.38-0.44 keV, as in the OverJet region, letting only the emission measure free to vary. Thanks to the new component, the quality of the fit drastically improved ($\chi^2$=613.77 with 506 d.o.f.). Since this new component showed oversolar abundances (Table \ref{risultatifit}), we associate it with shocked ejecta while the quite homogeneous soft x-ray emission in all the three regions is due to shocked ISM.   

A problem that emerged was the impossibility to estimate absolute abundances. Since spectra are dominated by line emission and the jet area is quite faint, it is impossible to estimate precisely the continuum and the intensity of each emission line: we can fit the data by enhancing all the abundances by the same factor $f$ and reducing the emission measure by $f$ without significant variation of $\chi^2$. In order to avoid partially this problem we considered relative abundances calculated with respect to Mg, the element which shows the most prominent line in the spectrum (see Fig. \ref{spettritot}). With this approach, we satisfyingly constrained abundances (provided that Mg abundance is fixed during the error calculation). Error on Mg abundance is calculated fixing the Fe abundance.   

\begin{figure}[!h]
\centering
\includegraphics[angle=270,width=\columnwidth]{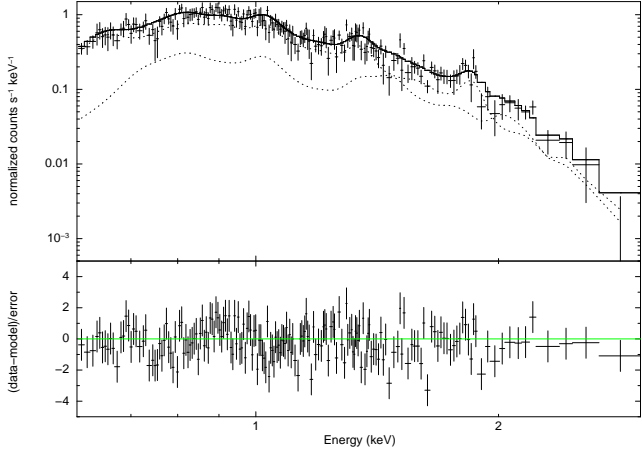}
\caption{EPIC PN spectrum of the jet extracted from observation 0114100501 showing two different components of the VAPEC+VRNEI model. In the lower panel residuals are displayed.}
\label{spettrojetsolopn}
\end{figure}

Since previous works (\citealt{out14}, \citealt{yok09}, M17) pointed out the presence of overionized plasma, we investigated whether the jet could show some indication of overionization. We substituted the hotter component of our two-components VAPEC model with a \emph{VRNEI} model, which is used to describe plasma in recombination phase because of its rapid cooling. To this end, we assumed an initial temperature of 5 keV (in agreement with M17). The $\chi^2$ significantly decreased from 613.77 (506 d.o.f ) to 575.05 (506 d.o.f.). This result suggests that ejecta in the jet are overionized. In Table \ref{risultatifit} we show the best fit results also for this VAPEC+VRNEI model for the jet. Fig. \ref{spettrojetsolopn} shows the PN-only spectrum extracted from observation 0114100501 with the contribution of the two spectral components. Below $\approx$ 1.4 keV, the ISM component is dominant while the VRNEI component becomes dominant at higher energies. In Fig. \ref{spettritot} spectra extracted from all cameras of both observations from regions OverJet and Jet are shown. It is clear that the spectrum of the region OverJet becomes significantly softer at higher energies than that of the jet region.

As it appears from Fig. \ref{dettagliojet}, the Front, namely the upper part of the jet, is different according to the energy band we are looking at, i.e. it is shifted to NE in the Hard band with respect to the Soft band. We analyzed spectra extracted from these regions finding that the Soft Front has chemical properties consistent within 2 $\sigma$ with those of the OverJet and UnderJet regions, being well described with a single VAPEC component with roughly solar abundances (Table \ref{spettrioverundersoft}). We interpret the enhanced brightness visible in Fig. \ref{dettagliojet} of this region as due to the interaction between the jet and the molecular cloud, as we will discuss in detail in Sect. \ref{discu}. On the other hand, the Hard Front's spectrum needs to be fitted with a two component VAPEC+VRNEI model providing a $\chi^2=346.33$ with 342 dof and showed abundances consistent with those of the jet within 2 $\sigma$ (Hard Front can also be described with a VAPEC+VAPEC model providing a $\chi^2$ = 342.57 with 342 dof). However, Hard Front plasma showed higher values of the ionization parameter ($\tau = 9 \pm 1 \cdot 10^{11}$ against $\tau = 3^{+2}_{-1} \cdot 10^{11}$ s/cm$^3$ found for the jet) and of the ratio Si/Mg $\approx 0.75$, which is  $\approx$ 10 times greater than in the jet (Si/Mg $\approx$ 0.08). This may be indicative of spatial inhomogeneities in physical and chemical properties of the jet. Nevertheless, the low surface brightness and poor statistics do not allow us to probe accurately this issue.

\begin{table}[!h]
\centering
\begin{tabular}{c|c|c}
\hline\hline
\multicolumn{3}{c}{Jet}\\
\hline
Parameter & \multicolumn{2}{c}{Model}\\
\hline
 & VAPEC+VAPEC& VAPEC+VRNEI\\
\hline
$nH (10^{22} cm^{-2})$ & 0.57 $\pm$ 0.03   & 0.60$^{+0.03}_{-0.01}$ \\
\hline
kT (keV)& 0.44& 0.38\\
$ n^2 l^{*}$ ($10^{18}$ cm$^{-2}$) &  5 $\pm 1$ & 4$\pm 3$ \\
\hline
 kT$_{init}$ (keV) & - & 5 (frozen) \\
 kT (keV) & 0.76$^{+0.08}_{-0.05}$& 0.21$^{+0.04}_{-0.03}$ \\
 O& 11$^{+7}_{-5}$ & 1 $(frozen)$ \\ 
 Ne&  4$^{+2}_{-3}$ & 11$^{+8}_{-5}$ \\
 Mg& 4$^{+1}_{-2} $& 13$^{+11}_{-7}$ \\
 Si& 0.8$^{+0.4}_{-0.6} $& 1$^{+2}_{-1}$ \\
 Fe& 0.8$\pm{0.3}$ & 10$^{+10}_{-6}$ \\
 Tau (s/cm$^3$) & - & 3$^{+2}_{-1} \cdot 10^{11}$ \\
 $ n^2 l^{*}$ ($10^{18}$ cm$^{-2}$) & 0.6$\pm 0.2$& 4.2$^{+0.9}_{-1.2}$ \\
\hline
$\chi^2$ (d.o.f)& 613.77 (506)&  575.05 (506)\\
 Counts & 20000 & \\
\hline\hline
\multicolumn{3}{c}{Hard Front}\\
\hline
$nH (10^{22} cm^{-2})$ & 0.73$\pm$0.05 & 0.99$^{+0.08}_{-0.10}$ \\
\hline
kT (keV) & 0.44& 0.38\\
$ n^2 l^{*}$ ($10^{18}$ cm$^{-2}$) & 4$^{+1}_{-2}$ & 4$^{+4}_{-3.9}$ \\ 
\hline
 kT$_{init}$ (keV) & - & 5 (frozen) \\
 kT (keV) &0.73$\pm$ 0.6 & 0.25 $\pm$ 0.03 \\
 O& 7$^{+8}_{-4}$& 1.6 $\pm$ 0.9 \\ 
 Ne& 2$^{+3}_{-2}$ & 3.6 $^{+1.3}_{-0.8}$ \\
 Mg& 3.1$^{+1.2}_{-1.4}$& 2.5 $\pm 1.2$ \\
 Si& 0.9$^{+0.7}_{-0.8}$ & 2.0$^{+0.8}_{-0.1}$ \\
 Fe& 1.8$^{+1.4}_{-0.6}$ & 2$^{+2}_{-1}$ \\
 Tau (s/cm$^3$) & - & 9$^{+2}_{1-} \cdot 10^{11}$ \\
 $ n^2 l^{*}$ ($10^{18}$ cm$^{-2}$) & 0.8$^{+0.5}_{-0.4}$& 1.3 $\pm$ 0.5 \\
\hline
$\chi^2$ (d.o.f)& 342.57 (342)&  346.33 (342)\\
 Counts & 11000 & \\
\hline
\multicolumn{3}{c}{$^*$ Emission measure per unit area}\\
\end{tabular}
\caption{Best fit results for the jet and Hard Front. S abundance is fixed equal to the Si one. Abundance of the first component are fixed to the best fit values displayed in Table \ref{spettrioverundersoft} and kT of the ISM component can vary only in the range 0.38-0.44 keV.}
\label{risultatifit}
\end{table}

\section{Discussion}
\label{discu}

In this paper we described our study of a Mg-rich jet-like structure detected in the NW area of IC443 whose emission is mainly due to overionized plasma. The jet projection towards the PWN crosses the position of the NS at the time of the explosion of the progenitor star, strongly indicating that the PWN belongs to IC443 and that the collimated jet has been produced by the exploding star.

The results of our spatially resolved spectral analysis helped us to imagine a reliable  scenario which can explain the observed morphology of the jet region (see Fig \ref{dettagliojet}.) The molecular cloud has an important role in the dynamical evolution of IC443. After the SN explosion, ejecta expanded and encountered the molecular cloud which has a density more than 100 times higher (\citealt{cck77}), representing a wall for the ejecta's plasma. In particular, the jet cannot pass through the cloud, is distorted and the distorted part is the region we called Hard Front which is the real upper part of the jet. The Soft Front emission is likely due to heating caused by the impact of the jet on the cloud. 

IC443's jet is the first structure of this type which clearly presents emission of overionized plasma. Until now other two jets have been inferred by \cite{wbv03}, \cite{hlb04} and \cite{fhm06} in Cas A and \cite{gsm17} in Vela SNR, but in both cases the plasma is underionized. We investigated possible causes of this overionization and we devised the following scenario. 

M17 studied the distribution of overionized plasma in IC443 by analyzing a set of Suzaku observations. They found overionized plasma in the region surrounding the PWN, where the ionization parameter is extremely low ( $ \tau \approx 4 \cdot 10^{11}$ s/cm$^3$), indicating a plasma very far from conditions of collisional ionization equilibrium (CIE). Instead, they found CIE conditions in the NW jet region (which is very distant from the SE cloud) and concluded that overionization is due to thermal conduction with the cold SE cloud. We here proved that, though a large area of the NW shell is in CIE (i.e. the part where the ISM emission dominates), the ejecta at NW are overionized. The ejecta in this region are concentrated in the collimated jet that is quite narrow (the thickest part is 1.5$'$ large) and therefore unresolved by the large PSF of the Suzaku mirrors. We also found that the degree of overionization in the NW jet ($\tau= 3^{+2}_{-1} \cdot 10^{11}$ s/cm$^3$) is perfectly consistent with that measured by M17 at SE ($\tau = 4.2^{+0.2}_{-0.1} \cdot{10^11}$ s/cm$^3$, see Table 3 of M17). This result seems to exclude that the overionization is due to thermal conduction with the SE cloud and may suggest an adiabatic cooling associated with the expansion of the ejecta. Before the expansion	, the ejecta may have been heated by the interaction with a reflected shock due to the impact of the forward shock front with the SE cloud (which is very close to the explosion site, marked by the original position of the PWN in Fig. \ref{ic443andjet}).

 Indeed, we roughly estimated temperature and ionization parameter of the plasma by assuming adiabatic expansion, starting with a temperature of 5 keV, which is the fixed value of kT$_{init}$ in our best fit with the VAPEC+VRNEI model (see Appendix A for a detailed description). We obtained $kT \approx 0.5$ keV and $ \tau \approx 2 \cdot 10^{11}$ s/cm$^3$ which are similar (within a factor 2) to those we measured with the spectral analysis (Table \ref{risultatifit}). These results further support our dynamical scenario (similar to that proposed for W49B by \citealt{zmb11}) which needs to be confirmed through dedicated hydrodynamic models. 

We estimated the mass and the kinetic energy associated with the jet and Hard Front using the best fit values. We obtained M$\approx 0.03$ M$_{sun}$ and K $\approx 4 \cdot 10^{48}$ erg which are intermediate values between those found by \cite{wbv03} and \cite{omp16} for Cas A (M $\approx 0.4$ M$_{sun}$ and K $\approx 10^{49}$ erg) and those found by \cite{gsm17} for the Vela SNR (M $\approx 0.008$ M$_{sun}$ and K $\approx 10^{47}$ erg). This similarity among all jets detected in core-collapse SNRs suggests that the physical origin of these structures is probably the same but details of the mechanism which cause them are still unknown. Moreover, we and \cite{gsm17} found that the jets cross the original position of the NS. It is, then, natural to speculate that this mechanism is somehow related to the core collapse of the progenitor star.

\begin{acknowledgements}
We appreciate the comments and suggestions by the referee. M.M., S.O. and G.P. acknowledge financial contribution from the agreement ASI-INAF n.2017-14-H.O.
\end{acknowledgements}

\appendix

\section{kT and $\tau$ estimates}

\begin{figure}[!h]
\centering
\includegraphics[width=\columnwidth]{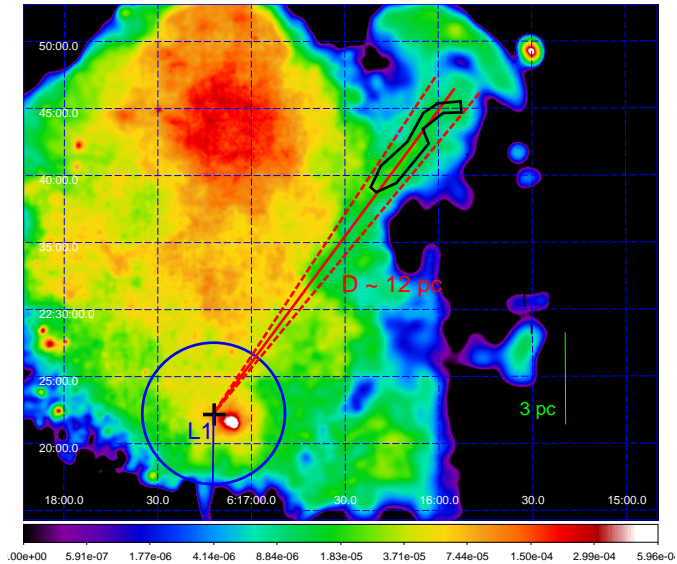}
\caption{Count-rate image of IC443 in the Hard energy band (1.4-5 keV). Overview of the scheme used to estimate the volume V1 of the jet at time $t_0$, namely when the forward shock interacts with the molecular cloud.}
\label{stimatauT}
\end{figure}

We roughly estimated temperature and ionization parameter of jet plasma with the following procedure in order to check whether our dynamical model for the plasma is reasonable. To get overionization one needs to have a hot plasma in CIE followed by a rapid cooling. For a gas in adiabatic expansion $T_1V_1^{\gamma-1}=T_2V_2^{\gamma-1}$ where, in our cases, $T_2$  and $V_2 \approx 10^{56}$ cm$^3$ are the current temperature and volume of the jet and $T_1 = 5$ keV, $V_1$ are the jet temperature and volume at the time in which the expanding forward shock interacts with the molecular cloud in the south (see Fig. \ref{stimatauT}). We assume that the ejecta are heated by the interaction with the molecular cloud in the south. If $L1$ $\approx$ 2 pc is the distance between the area of the explosion and the SE cloud and $v_{jet}$ is the velocity expansion (assumed constant) of the jet, then $t_0=L1/v_{jet}$ is the time when the hot ejecta are in CIE, immediately before their expansion. We then want to evaluate the volume of the jet at $t_0$ assuming it has a conic shape, as shown in Fig. \ref{stimatauT}. Considering the projection (marked with the letter D and long $\approx$ 12 pc) of the current jet (which has volume $V_2$) towards the PWN, it is possible to calculate $V_1 \approx 10^{54}$ cm$^3$ and then $T_2 \approx 0.5$ keV. 

The ionization parameter $\tau$ is defined as the time integral of the electron density $n_e$,  $ \tau= \int n_e dt$. $n_e$ is defined as $c/V$ where $c=m_{jet}/\mu_{prot}$ is the ratio between the jet mass and the average atomic mass $\mu_{prot}=2.1*10^{-24}$ g and $V$ is the volume occupied by the plasma. Thanks to the assumption of conic shape for the current jet it is possible to write $V=V_1+\alpha t^3$ where $\alpha=v_{jet}^3 \pi tg^2(\theta)/3$ which gives $\tau \approx 2 \cdot 10^{11}$ s/cm$^3$ and $kT_2 \approx 0.5$ keV. These values, similar to those measured through the spatially resolved spectral analysis, further confirm the goodness of our scenario.
    
\bibliographystyle{aa}

\begin{thebibliography}{30}
\expandafter\ifx\csname natexlab\endcsname\relax\def\natexlab#1{#1}\fi

\bibitem[{{Arnaud}(1996)}]{arn96}
{Arnaud}, K.~A. 1996, in ASP Conf. Ser. 101: Astronomical Data Analysis
  Software and Systems V, 17

\bibitem[{{Burton} {et~al.}(1988){Burton}, {Geballe}, {Brand}, \&
  {Webster}}]{bgb88}
{Burton}, M.~G., {Geballe}, T.~R., {Brand}, P. W. J.~L., \& {Webster}, A.~S.
  1988, \mnras, 231, 617

\bibitem[{{Cornett} {et~al.}(1977){Cornett}, {Chin}, \& {Knapp}}]{cck77}
{Cornett}, R.~H., {Chin}, G., \& {Knapp}, G.~R. 1977, \aap, 54, 889

\bibitem[{{Fesen} {et~al.}(2006){Fesen}, {Hammell}, {Morse}, {Chevalier},
  {Borkowski}, {Dopita}, {Gerardy}, {Lawrence}, {Raymond}, \& {van den
  Bergh}}]{fhm06}
{Fesen}, R.~A., {Hammell}, M.~C., {Morse}, J., {et~al.} 2006, \apj, 645, 283

\bibitem[{{Gaensler} {et~al.}(2006){Gaensler}, {Chatterjee}, {Slane}, {van der
  Swaluw}, {Camilo}, \& {Hughes}}]{gcs06}
{Gaensler}, B.~M., {Chatterjee}, S., {Slane}, P.~O., {et~al.} 2006, \apj, 648,
  1037

\bibitem[{{Garc{\'{\i}}a} {et~al.}(2017){Garc{\'{\i}}a}, {Su{\'a}rez},
  {Miceli}, {Bocchino}, {Combi}, {Orlando}, \& {Sasaki}}]{gsm17}
{Garc{\'{\i}}a}, F., {Su{\'a}rez}, A.~E., {Miceli}, M., {et~al.} 2017, \aap,
  604, L5

\bibitem[{{Grichener} \& {Soker}(2017)}]{gs17}
{Grichener}, A. \& {Soker}, N. 2017, \mnras, 468, 1226

\bibitem[{{Hwang} {et~al.}(2004){Hwang}, {Laming}, {Badenes}, {Berendse},
  {Blondin}, {Cioffi}, {DeLaney}, {Dewey}, {Fesen}, {Flanagan}, {Fryer},
  {Ghavamian}, {Hughes}, {Morse}, {Plucinsky}, {Petre}, {Pohl}, {Rudnick},
  {Sankrit}, {Slane}, {Smith}, {Vink}, \& {Warren}}]{hlb04}
{Hwang}, U., {Laming}, J.~M., {Badenes}, C., {et~al.} 2004, \apjl, 615, L117

\bibitem[{{Janka} {et~al.}(2016){Janka}, {Melson}, \& {Summa}}]{jms16}
{Janka}, H.-T., {Melson}, T., \& {Summa}, A. 2016, Annual Review of Nuclear and
  Particle Science, 66, 341

\bibitem[{{Kawasaki} {et~al.}(2002){Kawasaki}, {Ozaki}, {Nagase}, {Masai},
  {Ishida}, \& {Petre}}]{kon02}
{Kawasaki}, M.~T., {Ozaki}, M., {Nagase}, F., {et~al.} 2002, \apj, 572, 897

\bibitem[{{Matsumura} {et~al.}(2017){Matsumura}, {Tanaka}, {Uchida}, {Okon}, \&
  {Tsuru}}]{mtu17}
{Matsumura}, H., {Tanaka}, T., {Uchida}, H., {Okon}, H., \& {Tsuru}, T.~G.
  2017, \apj, 851, 73

\bibitem[{{Miceli} {et~al.}(2017){Miceli}, {Bamba}, {Orlando}, {Zhou},
  {Safi-Harb}, {Chen}, \& {Bocchino}}]{mbo17}
{Miceli}, M., {Bamba}, A., {Orlando}, S., {et~al.} 2017, \aap, 599, A45

\bibitem[{{Miceli} {et~al.}(2010){Miceli}, {Bocchino}, {Decourchelle},
  {Ballet}, \& {Reale}}]{mbd10}
{Miceli}, M., {Bocchino}, F., {Decourchelle}, A., {Ballet}, J., \& {Reale}, F.
  2010, \aap, 514, L2+

\bibitem[{{Ohnishi} {et~al.}(2014){Ohnishi}, {Uchida}, {Tsuru}, {Koyama},
  {Masai}, \& {Sawada}}]{out14}
{Ohnishi}, T., {Uchida}, H., {Tsuru}, T.~G., {et~al.} 2014, \apj, 784, 74

\bibitem[{{Orlando} {et~al.}(2016){Orlando}, {Miceli}, {Pumo}, \&
  {Bocchino}}]{omp16}
{Orlando}, S., {Miceli}, M., {Pumo}, M.~L., \& {Bocchino}, F. 2016, \apj, 822,
  22

\bibitem[{{Ozawa} {et~al.}(2009){Ozawa}, {Koyama}, {Yamaguchi}, {Masai}, \&
  {Tamagawa}}]{oky09}
{Ozawa}, M., {Koyama}, K., {Yamaguchi}, H., {Masai}, K., \& {Tamagawa}, T.
  2009, \apjl, 706, L71

\bibitem[{{Pavlov} {et~al.}(2003){Pavlov}, {Teter}, {Kargaltsev}, \&
  {Sanwal}}]{ptk03}
{Pavlov}, G.~G., {Teter}, M.~A., {Kargaltsev}, O., \& {Sanwal}, D. 2003, \apj,
  591, 1157

\bibitem[{{Rho} \& {Petre}(1998)}]{rp98}
{Rho}, J. \& {Petre}, R. 1998, \apjl, 503, L167

\bibitem[{{Sawada} \& {Koyama}(2012)}]{sk12}
{Sawada}, M. \& {Koyama}, K. 2012, \pasj, 64, 81

\bibitem[{{Smith}(2014)}]{smi14}
{Smith}, R. 2014, in COSPAR Meeting, Vol.~40, 40th COSPAR Scientific Assembly

\bibitem[{{Su} {et~al.}(2014){Su}, {Fang}, {Yang}, {Zhou}, \& {Chen}}]{sfy14}
{Su}, Y., {Fang}, M., {Yang}, J., {Zhou}, P., \& {Chen}, Y. 2014, \apj, 788,
  122

\bibitem[{{Swartz} {et~al.}(2015){Swartz}, {Pavlov}, {Clarke}, {Castelletti},
  {Zavlin}, {Bucciantini}, {Karovska}, {van der Horst}, {Yukita}, \&
  {Weisskopf}}]{spc15}
{Swartz}, D.~A., {Pavlov}, G.~G., {Clarke}, T., {et~al.} 2015, \apj, 808, 84

\bibitem[{{Troja} {et~al.}(2008){Troja}, {Bocchino}, {Miceli}, \&
  {Reale}}]{tbm08}
{Troja}, E., {Bocchino}, F., {Miceli}, M., \& {Reale}, F. 2008, \aap, 485, 777

\bibitem[{{Troja} {et~al.}(2006){Troja}, {Bocchino}, \& {Reale}}]{tbr06}
{Troja}, E., {Bocchino}, F., \& {Reale}, F. 2006, \apj, 649, 258

\bibitem[{{Uchida} {et~al.}(2012){Uchida}, {Koyama}, {Yamaguchi}, {Sawada},
  {Ohnishi}, {Tsuru}, {Tanaka}, {Yoshiike}, \& {Fukui}}]{uky12}
{Uchida}, H., {Koyama}, K., {Yamaguchi}, H., {et~al.} 2012, \pasj, 64, 141

\bibitem[{{Welsh} \& {Sallmen}(2003)}]{ws03}
{Welsh}, B.~Y. \& {Sallmen}, S. 2003, \aap, 408, 545

\bibitem[{{Willingale} {et~al.}(2003){Willingale}, {Bleeker}, {van der Heyden},
  \& {Kaastra}}]{wbv03}
{Willingale}, R., {Bleeker}, J.~A.~M., {van der Heyden}, K.~J., \& {Kaastra},
  J.~S. 2003, \aap, 398, 1021

\bibitem[{{Wilson}(1985)}]{wil85}
{Wilson}, J.~R. 1985, in Numerical Astrophysics, ed. J.~M. {Centrella}, J.~M.
  {Leblanc}, \& R.~L. {Bowers}, 422

\bibitem[{{Yamaguchi} {et~al.}(2009){Yamaguchi}, {Ozawa}, {Koyama}, {Masai},
  {Hiraga}, {Ozaki}, \& {Yonetoku}}]{yok09}
{Yamaguchi}, H., {Ozawa}, M., {Koyama}, K., {et~al.} 2009, \apjl, 705, L6

\bibitem[{{Zhou} {et~al.}(2011){Zhou}, {Miceli}, {Bocchino}, {Orlando}, \&
  {Chen}}]{zmb11}
{Zhou}, X., {Miceli}, M., {Bocchino}, F., {Orlando}, S., \& {Chen}, Y. 2011,
  \mnras, 415, 244

\end{thebibliography}

\end{document}